\documentclass[preprint,review,12pt]{elsarticle}
\usepackage{amsmath,amssymb,mathtools,amsthm}
\usepackage{bm,braket,booktabs,microtype}
\usepackage[hidelinks]{hyperref}
\hypersetup{pdfauthor={Ling-Zheng Xia, Lixin Xu},
 pdftitle={Proto-Area Response Beyond Bulk Entropy}}
\biboptions{sort&compress}
\newcommand{\eps}{\varepsilon}
\newcommand{\E}{\mathbb{E}}
\newcommand{\SPA}{S_{\mathrm{PA}}}
\newcommand{\Tr}{\operatorname{Tr}}
\newcommand{\Hess}{\operatorname{Hess}}
\newcommand{\cH}{\mathcal{H}}
\journal{Physics Letters B}
\begin{document}
\begin{frontmatter}
\title{Proto-Area Response Beyond Bulk Entropy}
\author[dlut]{Ling-Zheng Xia}
\author[dlut]{Lixin Xu\corref{cor1}}
\ead{lxxu@dlut.edu.cn}
\cortext[cor1]{Corresponding author.}
\affiliation[dlut]{organization={Institute of Theoretical Physics,
Dalian University of Technology},city={Dalian},postcode={116024},country={China}}

\begin{abstract}
We study how much of the proto-area response can be inferred from the
bulk von Neumann entropy. In the perturbative holographic code setting
considered by CCKLP and Witten, we keep the recovery frame fixed and
analyze the response averaged over the GUE at second order in the
encoding perturbation. Near the maximally mixed bulk spectrum, the
response is fixed by the entropy through quadratic order. For
$d_1\geq3$, spectral information first enters at cubic order through
the third centered moment, so states with the same entropy can
have different responses. We determine the resulting response range
on small equal-entropy level sets and obtain its sharp leading
$D^{3/2}$ dependence, including the coefficient. The corresponding
minimax error for prediction from entropy alone is one half of this
range. For $d_1=2$, the entropy instead determines the response
exactly.
\end{abstract}
\end{frontmatter}

\section{Introduction}

The Ryu--Takayanagi formula gives the relation between boundary
entanglement and bulk geometry a direct geometric form: the entropy
of a boundary region is determined, at leading order, by the area of
a bulk extremal surface \cite{RT:2006}. The quantum error-correcting
view of AdS/CFT helps explain why such an area term can coexist with
bulk information encoded redundantly in the boundary
\cite{ADH:2015,DHW:2016,HaPPY:2015,Hayden:2016,HarlowTASI:2018}.
For an exact subsystem code with a fixed code block, complementary
recovery makes the corresponding area contribution independent of
the logical state \cite{Beny:2007,Harlow:2017}. Once the encoding is
only approximate, this state independence need not survive. Cao, Cheng,
Karthikeyan, Li and Preskill (CCKLP) have
made this issue explicit by introducing proto-area entropy, defined
as the boundary entropy minus the entropy of the recovered bulk state
\cite{CCKLP:2026}. Witten has further studied a perturbative GUE deformation
of the encoding and obtained the leading state dependence in terms of
a modular spectral functional \cite{Witten:2026}.

An important feature of this functional is that it depends on the
full bulk entanglement spectrum, not just on the von Neumann entropy.
The distinction between entropy and finer spectral information is
also familiar from holographic R\'enyi entropy
\cite{Dong:2016,AkersRath:2019}. The question here is more specific:
whether the proto-area response itself can nevertheless be reduced,
near maximal mixing, to a function of the bulk entropy. Such a
reduction would require any two states with the same entropy to have
the same response. This is stronger than the monotonic relation seen
along a particular family of states. We therefore examine the order
at which spectral information independent of the entropy first enters,
and then ask how much the response can vary when the entropy is fixed
exactly.

We work with the response averaged over the GUE at second order in
the encoding perturbation, while keeping the recovery frame, the
short-distance state, and the dimensions fixed. For an arbitrary
full-rank short-distance Schmidt spectrum, the response near maximal
mixing is determined by the bulk entropy through quadratic order.
When $d_1\geq3$, a cubic spectral invariant remains after the entropy
contribution is removed. By constructing states with exactly equal
entropy, we show that this term rules out a local single-valued
relation between the response and the entropy. We then determine the
leading response range on a small entropy shell. The sharp
zero-sum third-moment bound \cite{Okumura:1974}, together with exact
entropy matching and uniform control of the remainder, gives an
explicit $D^{3/2}$ width and the associated minimax prediction error.

\section{Fixed-frame spectral response}
\label{sec:framework}
Let $p=(p_1,\ldots,p_{d_1})$ be the full-rank bulk spectrum and
$q=(q_1,\ldots,q_{d_2})$ the fixed full-rank short-distance Schmidt
spectrum, with $d_1\ge2$ and $d_2>1$. We perturb the encoding by
$e^{i\eps W}$, where $W$ is drawn from the GUE with variance
$\sigma_W^2$, and keep the exact-code decoder fixed.
The coefficient $B_\chi(p)$ measures the leading change in the mean
proto-area from its value at $p_{\rm mm}=(1/d_1,\ldots,1/d_1)$:
\begin{align}\label{eq:Bdef}
 \E\SPA^{\mathrm{fix}}(p)-\E\SPA^{\mathrm{fix}}(p_{\rm mm})
 &=\eps^2\sigma_W^2 B_\chi(p)+O(\eps^4).
\end{align}
Here proto-area is boundary entropy minus recovered bulk entropy,
and the superscript ``fix'' indicates that the recovery frame is
held fixed. Witten's
spectral calculation \cite{Witten:2026} gives
\begin{equation}\label{eq:Bchi}
 B_\chi(p)=-d_2^2[\Xi_A(p)-\Xi_A(p_{\rm mm})]+\Xi(p)-d_1^2.
\end{equation}
The two terms include the boundary-entropy response and the
finite-dimensional recovered entropy subtraction. In particular,
$B_\chi(p_{\rm mm})=0$.

The functionals are built from Witten's even kernel. With
$k_i=-\log p_i$ and $\ell_\mu=-\log(d_2q_\mu)$,
\begin{equation}\label{eq:Xispectral}
 L(x)=x\coth(x/2),\quad L(0)=2,\qquad
 \Xi(p)=\frac12\sum_{ij}L(k_i-k_j),
\end{equation}
\begin{equation}\label{eq:XiA}
 \Xi_A(p;\chi)=\frac1{2d_2^2}\sum_{ij\mu\nu}
 L(k_i-k_j+\ell_\mu-\ell_\nu).
\end{equation}
For flat $q$, $\Xi_A=\Xi$ and
$B_{\rm flat}=-(d_2^2-1)(\Xi-d_1^2)$.
\ref{app:encoding} gives the encoding conventions and
relative-entropy representation.

\section{Entropy-only description and its first obstruction}
\label{sec:quadratic}
To what order does bulk entropy determine $B_\chi$? Near maximal
mixing it determines the quadratic response in every spectral direction,
but generally fails at cubic order. Let
$S_a(p)=-\sum_i p_i\log p_i$ and
$\Delta S_a(p)=\log d_1-S_a(p)$.
Write $p=p_{\rm mm}+\delta p$ with $\delta p$ in the spectral tangent
space $\mathcal T=\{v\in\mathbb R^{d_1}:\sum_i v_i=0\}$, and let
$I_{\mathcal T}$ denote the identity on $\mathcal T$. All norms are Euclidean,
and all Hessians in this section are evaluated at $p_{\rm mm}$
and restricted to $\mathcal T$. Throughout this section, the next
section, and their proofs, $d_1$, $d_2$, and the short-distance state
$\chi$ are fixed, with a full-rank reduction on $A_2$.
Constants in $O(\cdot)$ may depend on these fixed quantities.
The coefficient governing the proto-area/entropy relation is
\begin{equation}\label{eq:cchi-hessian}
 \Gamma_\chi=\sum_{\mu,\nu=1}^{d_2}
 L''(\ell_\mu-\ell_\nu)-\frac13.
\end{equation}

Our first result is that entropy determines the quadratic response.
At $p_{\rm mm}$ the first variations of $B_\chi$ and $S_a$ vanish,
and their Hessians satisfy
\begin{equation}\label{eq:response}
 \Hess B_\chi=d_1^2\Gamma_\chi\Hess S_a
             =-d_1^3\Gamma_\chi I_{\mathcal T},
 \qquad \Gamma_\chi>0.
\end{equation}
Thus, in every bulk spectral direction,
\begin{equation}\label{eq:quadratic-prediction}
 B_\chi(p)=-d_1^2\Gamma_\chi\Delta S_a(p)
           +O(\|\delta p\|^3).
\end{equation}

Permutation symmetry and the trace constraint leave a single
quadratic invariant on $\mathcal T$:
$\sum_{ij}(v_i-v_j)^2=2d_1\sum_i v_i^2$.
The kernel curvature fixes its coefficient. \ref{app:spectral} gives the spectral expansion and
$\Gamma_\chi$ bounds. The underlying kernel bounds are proven in \ref{app:bounds-regime}.

At second order in the encoding and through quadratic order in the
bulk spectral deviation, the coefficient relating the mean proto-area
change to $S_a(p)-S_a(p_{\rm mm})=-\Delta S_a(p)$ is
$\eps^2\sigma_W^2d_1^2\Gamma_\chi$. For a flat short-distance
spectrum, $\Gamma_\chi=(d_2^2-1)/3$, recovering
$\eps^2\sigma_W^2d_1^2(d_2^2-1)/3$. The equality condition in Eq.~\eqref{eq:Gamma-bound} implies that every
nonflat short-distance spectrum reduces this coefficient.

To determine where this description fails, subtract the quadratic
entropy prediction from $B_\chi(p)$. The first correction is
\begin{equation}\label{eq:unified-series}
 B_\chi(p)+d_1^2\Gamma_\chi\Delta S_a(p)=\frac{d_1^4\Gamma_\chi}{3}\sum_i\delta p_i^3
           +O(\|\delta p\|^4).
\end{equation}
The cubic residual is controlled by the third spectral moment
$\sum_i\delta p_i^3$.

A nonzero residual alone would only mean that the quadratic
entropy prediction needs correction. To rule out any single-valued relation
$B_\chi=F(S_a)$, we compare states with exactly the same entropy.
For $d_1\ge3$ take
\begin{align}\label{eq:paths}
 p^{(1)}(t)&=p_{\rm mm}+t(1,-1,0,\ldots,0),\nonumber\\
 p^{(2)}(s)&=p_{\rm mm}+\frac{s}{\sqrt3}(1,1,-2,0,\ldots,0).
\end{align}
As proved in \ref{app:equal}, an analytic reparametrization satisfies
\begin{equation}\label{eq:reparam}
 s(t)=t-\frac{d_1}{6\sqrt3}t^2+O(t^3),\qquad
 S_a(p^{(2)}(s(t)))=S_a(p^{(1)}(t)).
\end{equation}
The entropy terms in Eq.~\eqref{eq:unified-series} cancel exactly,
leaving
\begin{equation}\label{eq:no-go}
 B_\chi(p^{(2)}(s(t)))-B_\chi(p^{(1)}(t))
 =-\frac{2d_1^4\Gamma_\chi}{3\sqrt3}t^3+O(t^4).
\end{equation}
Since $\Gamma_\chi>0$, no function of entropy alone agrees with
$B_\chi$ throughout any neighbourhood of maximal mixing. The
finite-dimensional recovered entropy subtraction does not cancel
this obstruction. For $d_1=2$, entropy fixes the spectrum up to
permutation, so the permutation-invariant $B_\chi$ admits an
entropy-only description on the full-rank binary simplex.
\ref{app:binary} gives an exact check.

The quadratic validity and cubic obstruction refer to orders in
the bulk spectral deviation, while $B_\chi$ remains a second-order
encoding coefficient.

\section{Sharp equal-entropy ambiguity}
\label{sec:width}
Section~\ref{sec:quadratic} shows that equal-entropy bulk states
can produce different leading changes in the mean fixed-frame
proto-area. This ambiguity is the range of $B_\chi$ at a fixed
entropy deficit. For $d_1\ge3$, fix a sufficiently small entropy
deficit $D>0$ and define
\begin{align}\label{eq:width-def}
 &\mathcal P_D
 =\{p_i>0:\ \sum_i p_i=1,\ \Delta S_a(p)=D\},\nonumber\\
 &B_{\chi,+}(D)=\max_{p\in\mathcal P_D}B_\chi(p),\qquad
 B_{\chi,-}(D)=\min_{p\in\mathcal P_D}B_\chi(p),\nonumber\\
 &\mathcal W_\chi(D)=B_{\chi,+}(D)-B_{\chi,-}(D).
\end{align}
This level set is compact and bounded away from the boundary
of the simplex. The width
$\mathcal W_\chi(D)$ measures the uncertainty in the response
coefficient when the bulk entropy alone is known. Its contribution
to the mean proto-area width at second encoding order is
$\eps^2\sigma_W^2\mathcal W_\chi(D)$.

Our second result gives the size of this ambiguity:
\begin{equation}\label{eq:width-law}
 \mathcal W_\chi(D)
 =\frac{2^{5/2}d_1^2(d_1-2)}{3\sqrt{d_1-1}}\Gamma_\chi D^{3/2}
  +O(D^2),\qquad D\to0^+.
\end{equation}
Writing $p=p_{\rm mm}+r(D,v)v$, let $r(D,v)>0$ be fixed by
the exact entropy constraint. Spectra along the unit directions
\begin{equation}\label{eq:edge-directions}
 v_+=\frac{(d_1-1,-1,\ldots,-1)}{\sqrt{d_1(d_1-1)}},
 \qquad v_-=-v_+,
\end{equation}
reach the largest and smallest responses to leading order,
respectively. The remainder is uniform over bulk spectral directions.

The extremization reduces to the sharp zero-sum third-moment
bound \cite{Okumura:1974}. Exact entropy matching and a remainder
uniform over all spectral directions make the resulting width
sharp. \ref{app:width} gives the proof.

For a sufficiently small fixed entropy deficit $D>0$, the continuous
radial parametrization identifies $\mathcal P_D$ with the unit sphere
in $\mathcal T$, which is connected for $d_1\ge3$.
The response therefore varies over the interval
$[B_{\chi,-}(D),B_{\chi,+}(D)]$ as the bulk spectrum varies over
$\mathcal P_D$. An entropy-only predictor cannot distinguish spectra within this set,
so it must assign a single value $z(D)$. The optimal choice is
the midpoint of the interval, giving the minimax error
\begin{equation}\label{eq:minimax}
 \inf_{z\in\mathbb R}\sup_{p\in\mathcal P_D}|B_\chi(p)-z|
 =\frac12\mathcal W_\chi(D).
\end{equation}
The quadratic predictor
$-d_1^2\Gamma_\chi D$ has the same leading error because the
two edges are symmetric about it through that order.

The width quantifies the deterministic ambiguity due to spectral
information that bulk entropy leaves undetermined.
Multiplying Eq.~\eqref{eq:minimax} by $\eps^2\sigma_W^2$ gives the
prediction limit for the second-order mean proto-area change.
The full shell width relative to the quadratic response magnitude is
\begin{equation}\label{eq:relative-width}
 \frac{\mathcal W_\chi(D)}{d_1^2\Gamma_\chi D}
 =\frac{2^{5/2}(d_1-2)}{3\sqrt{d_1-1}}\sqrt D+O(D).
\end{equation}
The short-distance spectrum changes the absolute width and response
together, and cancels from their leading ratio.

\section{Discussion}

Near maximal mixing, the bulk entropy contains all the spectral
information needed for the response through quadratic order. This
ceases to be true at cubic order when $d_1\geq3$. The surviving
spectral invariant distinguishes states that have the same
entropy, and the response range on a small entropy shell is therefore
nonzero. Its leading behavior is proportional to $D^{3/2}$, with the
coefficient given in Eq.~\eqref{eq:width-law}. The smallest possible
worst-case error for a prediction based only on the entropy is one
half of this range. The ambiguity discussed here is thus a property
of the spectrum after the GUE average has already been performed.

The appearance of the quadratic and cubic terms has a simple
information geometric interpretation. On the commuting probability
simplex, permutation symmetry and the trace constraint leave only one
quadratic invariant at maximal mixing, which is proportional to the
classical Fisher metric. At cubic order, an additional symmetric
invariant is available for $d_1\geq3$, namely $\sum_i v_i^3$.
With
$T_p(u,v,w)=\sum_i{u_i v_i w_i}/{p_i^2}$,
this is the contraction of the Amari--Chentsov tensor,
$T_{p_{\rm mm}}(v,v,v)=d_1^2\sum_i v_i^3$
\cite{AmariNagaoka:2000}. The modular kernel fixes the coefficients
of these invariants through $\Gamma_\chi$. Exact entropy matching is
then what turns the local cubic term into a genuine difference between
states on the same entropy shell.

The relative-entropy expansion is governed by the BKM metric. On
commuting states it reduces to classical Fisher geometry, whereas in
general it differs from the SLD/Bures metric
\cite{Petz:1996,LesniewskiRuskai:1999,BraunsteinCaves:1994}.
There is a related gravitational structure in holography: for
perturbations of the vacuum reduced to a ball-shaped region, the
second variation of relative entropy is related to canonical energy
in the corresponding AdS-Rindler wedge \cite{LashkariVR:2016}.
The quantum correction to the RT formula \cite{FLM:2013}, the JLMS
relation \cite{JLMS:2016}, and the entanglement first law
\cite{Blanco:2013} supply the broader semiclassical setting for this
connection. Entanglement first-law arguments for all ball-shaped
regions are also known to reproduce the linearized Einstein equations
\cite{Faulkner:2014,Lashkari:2014}. However, none of these results identifies 
the spectral width $\mathcal W_\chi$ found here
with a width of geometric areas. Such an interpretation would require
a further gravitational dictionary.

Our calculation concerns the GUE-averaged response at second order in
the encoding perturbation, with the recovery frame, finite dimensions,
and full-rank short-distance spectrum $q$ fixed. The estimates need
not be uniform as dimensions grow or $q$ approaches the boundary of
the simplex. Extension to optimized recovery remains
open \cite{Witten:2026,Cotler:2019,Chen:2020}. A recovery correction that is
small relative to the total response can still alter the finer
$D^{3/2}$ separation. Extending this sharp prediction limit beyond a
fixed recovery frame requires uniform control over how recovery
corrections vary across states of equal entropy.

\section*{Acknowledgements}
This work was supported by the National Natural Science Foundation of
China under Grant No.~12475047.

\section*{Data availability}
No data were used for the research described in the article.

\appendix
\section{Fixed-frame response and spectral expansion}
\label{app:kernel}
\subsection{Encoding and relative-entropy representation}
\label{app:encoding}
For a boundary region $A$ and its complement $\bar A$, let the
CFT Hilbert space factorize as $\cH=\cH_A\otimes\cH_{\bar A}$ with
$\cH_A=\cH_{A_1}\otimes\cH_{A_2}$ and
$\cH_{\bar A}=\cH_{\bar A_1}\otimes\cH_{\bar A_2}$, where the first
factor of each region is the low-energy sector and the second the
short-distance sector \cite{Witten:2026}. Set
$d_1=\dim\cH_{A_1}=\dim\cH_{\bar A_1}\ge2$ and
$d_2=\dim\cH_{A_2}=\dim\cH_{\bar A_2}>1$. The unperturbed code is
exact: in the fixed recovery frame (the identity decoder, with no
recovery unitaries), the encoding is the isometry
$V_0:\cH_{A_1}\otimes\cH_{\bar A_1}\to\cH$ given by
$V_0\ket\psi=\ket\psi\otimes\ket\chi$, where the low-energy bulk
state $\psi\in\cH_{A_1}\otimes\cH_{\bar A_1}$ and the short-distance
state $\chi\in\cH_{A_2}\otimes\cH_{\bar A_2}$ are pure and normalized.
Their reductions
$\sigma_{A_1}^{(0)}=\Tr_{\bar A_1}|\psi\rangle\langle\psi|$ and
$\sigma_{A_2}^{(0)}=\Tr_{\bar A_2}|\chi\rangle\langle\chi|$ are full
rank, with spectra $p_i>0$ and $q_\mu>0$. The perturbed encoding is $V_\eps=e^{i\eps W}V_0$, where $\eps$ is
a real parameter and $W$ a Hermitian matrix drawn from the Gaussian
unitary ensemble (GUE), with second moment
\begin{equation}\label{eq:gauss}
 \E[W_{IJ}W_{KL}]=\sigma_W^2\delta_{IL}\delta_{JK},
\end{equation}
where the capital indices run over an orthonormal basis of $\cH$.
Here $\E$ denotes the average over the GUE and $\sigma_W$ sets the
perturbation strength.
All spectral variations below keep the ensemble, the short-distance
state, and this decoder fixed. For $X\in\{A,A_1,A_2\}$ write
$\sigma_X^{(\eps)}=\Tr_{X^c}|V_\eps\psi\rangle\langle V_\eps\psi|$,
where $X^c$ is the complement of $X$. Then
$\sigma_A^{(0)}=\sigma_{A_1}^{(0)}\otimes\sigma_{A_2}^{(0)}
=\operatorname{diag}(p)\otimes\operatorname{diag}(q)$.
With $S(\rho)=-\Tr\rho\log\rho$, the fixed-frame proto-area entropy
is
\begin{equation}\label{eq:SPA}
 \SPA^{\mathrm{fix}}=S(\sigma_A^{(\eps)})-S(\sigma_{A_1}^{(\eps)}).
\end{equation}
CCKLP's definition involves an optimization over recovery channels. Witten's
analysis of the recovery correction \cite{Witten:2026} motivates the
fixed-frame calculation used below.

For $\rho,\sigma$ on the same Hilbert space, the relative entropy is 
$D(\rho\Vert\sigma)=\Tr\rho(\log\rho-\log\sigma)$. The fixed-frame 
proto-area entropy can be written as
\begin{equation}\label{eq:SPArel}
 \SPA^{\mathrm{fix}}=-D_A+D_{A_1}+Y,
 \qquad Y=-\Tr\sigma_{A_2}^{(\eps)}\log\sigma_{A_2}^{(0)},
\end{equation}
where $D_X=D(\sigma_X^{(\eps)}\Vert\sigma_X^{(0)})$.
For a maximally mixed short-distance reduction, $q_\mu=1/d_2$ for all
$\mu$ ($\chi$ is then maximally entangled). In this case, $Y$ equals
$\log d_2$, the entropy $S_\chi$ of the short-distance reduction.
The second-order Kubo--Mori expansion of the relative entropy is
$D(\sigma+\delta\Vert\sigma)=\frac12\sum_{u,u'}
w(r_u,r_{u'})|\delta_{uu'}|^2+O(\delta^3)$,
where $r_u$ are the eigenvalues of $\sigma$ and $\delta$ is traceless \cite{Kubo:1957}.
In this flat-$q$ case, it gives \cite{Witten:2026}
\begin{align}\label{eq:expectations}
 \E D_{A_1}&=\eps^2\sigma_W^2(\Xi-1)+O(\eps^4),\nonumber\\
 \E D_A&=\eps^2\sigma_W^2(d_2^2\Xi-1)+O(\eps^4),
\end{align}
with
\begin{align}\label{eq:Xidef}
 \Xi(p)&=\frac12\sum_{i,j=1}^{d_1}(p_i+p_j)w(p_i,p_j),\nonumber\\
 w(p,p')&=\frac{\log p-\log p'}{p-p'},\qquad w(p,p)=\frac1p.
\end{align}
The kernel $w$ is the Bogoliubov--Kubo--Mori (BKM) metric kernel
\cite{Petz:1996,LesniewskiRuskai:1999}. The even modular kernel $L$ has the local expansion
\begin{equation}\label{eq:Lexpansion}
 L(x)=2+\frac{x^2}{6}-\frac{x^4}{360}+O(x^6).
\end{equation}

The dimensionless coupling
\begin{equation}\label{eq:lambda}
 \lambda=\eps^2\sigma_W^2d_1^2d_2^2
\end{equation}
is the mean squared size of the perturbation $\eps W$ on a
normalized vector.

For general $q$, Eq.~\eqref{eq:XiA} follows by using the product
eigenvalues $p_iq_\mu$ in the same expansion.
In Eq.~\eqref{eq:expectations}, $\Xi$ is replaced by $\Xi_A$
for $D_A$, while $D_{A_1}$ is unchanged. The GUE second moment also gives
$\E\sigma_{A_2}^{(\eps)}=\sigma_{A_2}^{(0)}
+\lambda(I_{A_2}/d_2-\sigma_{A_2}^{(0)})+O(\eps^4)$. Thus the
mean of $Y$ is independent of $p$ through this order. Subtracting
the value at $p_{\rm mm}$ in Eq.~\eqref{eq:SPArel} gives
Eq.~\eqref{eq:Bchi}, including the recovered entropy subtraction.

\subsection{Spectral expansion and \texorpdfstring{$\Gamma_\chi$}{Gamma-chi}}
\label{app:spectral}

To compute the Hessian of $B_\chi$, use its expression in terms of
$\Xi_A$ and $\Xi$ in Eq.~\eqref{eq:Bchi}. The function
$G_\chi(x)=d_2^{-2}\sum_{\mu\nu}L(x+\ell_\mu-\ell_\nu)$
is even: $L(-x)=L(x)$ and exchanging $\mu\leftrightarrow\nu$
gives $G_\chi(-x)=G_\chi(x)$. Thus $G_\chi'(0)=0$ and
$c_\chi:=G_\chi''(0)=(\Gamma_\chi+1/3)/d_2^2$.
We use this curvature to calculate the two spectral Hessians. Since
$k_i-k_j=-d_1(\delta p_i-\delta p_j)+O(\|\delta p\|^2)$ and
$\sum_{ij}(v_i-v_j)^2=2d_1\sum_i v_i^2$ on $\mathcal T$,
\begin{equation}\label{eq:hessian}
 \Hess\Xi_A=d_1^3c_\chi I_{\mathcal T},\qquad
 \Hess\Xi=\frac{d_1^3}{3}I_{\mathcal T}.
\end{equation}
Their first variations vanish. Combining these Hessians in
Eq.~\eqref{eq:Bchi} and using $\Hess S_a=-d_1I_{\mathcal T}$
gives Eq.~\eqref{eq:response}.

The kernel bounds proved in \ref{app:bounds-regime} imply
\begin{equation}\label{eq:Gamma-bound}
 \frac{d_2-1}{3}<\Gamma_\chi\le\frac{d_2^2-1}{3},
\end{equation}
with equality on the right only for a flat short-distance
spectrum. Thus $\Gamma_\chi>0$. Finally, Taylor expansion of
$B_\chi$ and $S_a$, whose first variations vanish, proves
Eq.~\eqref{eq:quadratic-prediction}.

Expanding to cubic order gives
\begin{align}\label{eq:local-series}
 \Delta S_a(p)&=\frac{d_1}{2}\sum_i\delta p_i^2
 -\frac{d_1^2}{6}\sum_i\delta p_i^3+O(\|\delta p\|^4),\nonumber\\
 \Xi_A(p)-\Xi_A(p_{\rm mm})&=\frac{d_1^3c_\chi}{2}\sum_i\delta p_i^2
 -\frac{d_1^4c_\chi}{2}\sum_i\delta p_i^3
 +O(\|\delta p\|^4).
\end{align}
Here the cubic gap sum follows from
$\sum_{ij}(\delta p_i-\delta p_j)(\delta p_i^2-\delta p_j^2)
=2d_1\sum_i\delta p_i^3$. The expansion of $\Xi-d_1^2$ follows by
setting $c_\chi=1/3$. Substituting into Eq.~\eqref{eq:Bchi} gives Eq.~\eqref{eq:unified-series}.

\subsection{Kernel bounds and perturbative regime}
\label{app:bounds-regime}
The local expansion \eqref{eq:Lexpansion} is the beginning of the Bernoulli series
$L(x)=2+2\sum_{n\ge1}B_{2n}x^{2n}/(2n)!$ with $B_2=1/6$ and
$B_4=-1/30$, convergent for $|x|<2\pi$. Set $u=x/2$ and
$g(u)=L''(2u)=(u\coth u-1)/\sinh^2u$.
For $u>0$,
\begin{equation}
 g'(u)=\frac{h(u)}{2\sinh^4u},\qquad
 h(u)=3\sinh(2u)-4u-2u\cosh(2u).
\end{equation}
Here $h(0)=h'(0)=0$ and
$h''(u)=4[\sinh(2u)-2u\cosh(2u)]<0$ for $u>0$,
because $\tanh(2u)<2u$. Thus $L''$ strictly decreases on the
positive axis. It is even, $L''(0)=1/3$, and
$L''(x)\sim2xe^{-x}\to0$ as $x\to+\infty$, proving
$0<L''(x)\le1/3$ and the stated equality condition.

The $d_2$ diagonal terms in $\Gamma_\chi+1/3$ contribute
$d_2/3$, and every off-diagonal term is strictly positive and at
most $1/3$. This proves Eq.~\eqref{eq:Gamma-bound}. The upper
bound is reached only when all modular gaps vanish, that is, for a
flat short-distance spectrum. The lower endpoint is approached
when all off-diagonal modular gaps grow without bound.

The remainders in Eq.~\eqref{eq:expectations} refer to
$\eps\to0$ with dimensions and full-rank spectra fixed. Ensemble
symmetry removes the odd powers. We extract the encoding coefficient
before taking $D\to0$. At finite encoding strength, a conservative sufficient
condition for resolving the cubic spectral signal against uncontrolled
higher-order encoding corrections is
\begin{equation}\label{eq:window}
 \Gamma_\chi D^{3/2}\gg\lambda C,
\end{equation}
where $\lambda=\eps^2\sigma_W^2d_1^2d_2^2$ and $C$ is a finite
constant at fixed dimensions and fixed full-rank spectra. The spectral
remainder estimates are controlled at fixed
finite dimensions and for a fixed full-rank short-distance spectrum
$q$, and remain uniform over bulk spectral directions in the
small-deficit regime considered here. Extending these estimates
uniformly to growing dimensions or to short-distance spectra
approaching the simplex boundary would require separate control of
the corresponding constants.

\section{Exact entropy matching}\label{app:equal}
For the paths in Eq.~\eqref{eq:paths}, define
$f_a(t)=\Delta S_a(p^{(a)}(t))$. Their expansions are
\begin{equation}
 f_1(t)=d_1t^2+O(t^4),\qquad
 f_2(t)=d_1t^2+\frac{d_1^2}{3\sqrt3}t^3+O(t^4).
\end{equation}
The implicit-function theorem cannot be applied directly to
$f_2(s)-f_1(t)$ at the origin, where both first derivatives vanish.
Instead put $s=tz$ and extend
$H(t,z)=[f_2(tz)-f_1(t)]/t^2$ analytically to $t=0$.
Then $H(0,z)=d_1(z^2-1)$ and
$\partial_zH(0,1)=2d_1\ne0$. The implicit-function theorem gives
an analytic $z(t)$ near $1$, and hence the exactly entropy-matching
$s(t)=tz(t)$ in Eq.~\eqref{eq:reparam}. Substitution into
Eq.~\eqref{eq:unified-series} proves Eq.~\eqref{eq:no-go}.

\section{Exact binary check}\label{app:binary}
For binary spectra the conclusion is different.
For $d_1=2$, write the spectrum as $(p,1-p)$, with $0<p<1$,
and set $x=\log[p/(1-p)]$. Writing $B_\chi(x)$ for its response,
Eq.~\eqref{eq:Bchi} gives exactly
\begin{equation}
 B_\chi(x)=-d_2^2[G_\chi(x)-G_\chi(0)]+L(x)-2.
\end{equation}
Here $G_\chi(x)=d_2^{-2}\sum_{\mu\nu}L(x+\ell_\mu-\ell_\nu)$.
Both kernels are even, so $B_\chi(x)=B_\chi(-x)$. The binary
entropy is also even in $x$ and has derivative
$-x/[4\cosh^2(x/2)]<0$ for $x>0$. It therefore determines
$|x|$ uniquely, and hence determines $B_\chi$ exactly.

\section{Sharp third-moment bound and uniform entropy shell}
\label{app:width}
For $d_1\ge3$, let $\mathbb S_{\mathcal T}$ be the compact unit sphere
in $\mathcal T$. At an extremum of $\sum_i v_i^3$ subject to
$\sum_i v_i=0$ and $\sum_i v_i^2=1$, the two constraint gradients
are independent. Lagrange multipliers give
$3v_i^2=\alpha+2\beta v_i$, so there are at most two component
values. If the positive value $a$ has multiplicity $k$ and the
negative value $b$ multiplicity $d_1-k$, the constraints give
\begin{equation}
 a=\sqrt{\frac{d_1-k}{d_1k}},\qquad
 b=-\sqrt{\frac{k}{d_1(d_1-k)}},\qquad
 \sum_i v_i^3=\frac{d_1-2k}{\sqrt{d_1k(d_1-k)}}.
\end{equation}
Its squared value is $d_1/[k(d_1-k)]-4/d_1$, maximized at $k=1$ or
$d_1-1$. This proves
\begin{equation}\label{eq:moment-bound}
 \left|\sum_i v_i^3\right|
 \le\frac{d_1-2}{\sqrt{d_1(d_1-1)}},
\end{equation}
with equality at $v_\pm$ and their permutations. This is the
zero-sum form of the Okumura inequality
\cite{Okumura:1974}.

On the closed probability simplex, entropy extends continuously
and has a unique maximum at $p_{\rm mm}$. On the compact complement
of any neighbourhood of that point, its deficit has a strictly
positive minimum. Consequently every sufficiently small positive
entropy shell lies in a ball $\|p-p_{\rm mm}\|<\rho$, where
$\rho<1/(2d_1)$ may be fixed in advance. All probabilities there
are bounded below by $1/(2d_1)$. The shell is closed in the full
simplex and thus compact. The radial entropy deficit is strictly
increasing for $r>0$ in this ball: its second derivative is
$\sum_i v_i^2/(1/d_1+rv_i)>0$ and its first derivative vanishes at
$r=0$. Every direction therefore intersects a sufficiently small
shell exactly once.

For $v\in\mathbb S_{\mathcal T}$, put
$F(r,v)=\Delta S_a(p_{\rm mm}+rv)$. On this common ball Taylor's
theorem gives
\begin{equation}\label{eq:uniform-entropy}
 F(r,v)=\frac{d_1}{2}r^2-\frac{d_1^2}{6}\Bigl(\sum_i v_i^3\Bigr)r^3
       +O(r^4),
\end{equation}
with a constant independent of $v$, since fourth derivatives are
bounded on its compact closure. The function
$h(r,v)=\sqrt{2F(r,v)/(d_1r^2)}$ extends smoothly through $r=0$
with $h(0,v)=1$. Let $u=\sqrt{2D/d_1}$. Applying the implicit-function
theorem to $rh(r,v)=u$, uniformly on the compact sphere, yields
\begin{equation}\label{eq:uniform-radius}
 r(D,v)=u+\frac{d_1}{6}\Bigl(\sum_i v_i^3\Bigr)u^2+O(u^3).
\end{equation}
For each sufficiently small fixed $D>0$, the radius $r(D,v)$ is
smooth in $v$, so $v\mapsto p_{\rm mm}+r(D,v)v$ is a continuous
parametrization of $\mathcal P_D$ with continuous inverse
$p\mapsto(p-p_{\rm mm})/\|p-p_{\rm mm}\|$.
In particular, $r^3=u^3+O(u^4)$ uniformly.

Substituting the uniform radius expansion into
Eq.~\eqref{eq:unified-series} gives
\begin{equation}\label{eq:directional-width}
 B_\chi(p)=-d_1^2\Gamma_\chi D
 +\frac{2^{3/2}d_1^{5/2}\Gamma_\chi}{3}
 D^{3/2}\sum_i v_i^3+O(D^2).
\end{equation}

At fixed $\chi$ with full-rank reduction, the fourth bulk derivatives of $B_\chi$
are bounded on the same ball. The remainder in
Eq.~\eqref{eq:unified-series} therefore satisfies, for some finite
$C>0$,
\begin{equation}\label{eq:uniform-remainder}
 \left|B_\chi(p_{\rm mm}+rv)+d_1^2\Gamma_\chi F(r,v)
 -\frac{d_1^4\Gamma_\chi}{3}r^3\sum_i v_i^3\right|
 \le Cr^4
\end{equation}
for every $v\in\mathbb S_{\mathcal T}$ and $0\le r\le\rho$.
Equations~\eqref{eq:uniform-radius} and~\eqref{eq:uniform-remainder}
prove the uniform $O(D^2)$ remainder in
Eq.~\eqref{eq:directional-width}. We can therefore take the maximum
and minimum over all directions. The values at $v_+$ and $v_-$
reach the bounds through order $D^{3/2}$, so the width formula
follows.

\bibliographystyle{elsarticle-num}
\bibliography{references}
\end{document}